\documentclass[aps,twocolumn,showlabels,showrefs,amsmath,amssymb,pre,superscriptaddress,floatfix,colors]{revtex4-1}

\usepackage{graphicx}
\usepackage{dcolumn}
\usepackage{bm}
\usepackage{amssymb}
\usepackage{multirow}
\usepackage{color}
\usepackage{hyperref}
\usepackage{ulem}
\usepackage{float}

\begin{document}

\title{A particle-based Ising model.}

\date{\today}


\author{Quentin Novinger}
\affiliation{Institute for Computational Molecular Science, Temple University, Philadelphia, PA 19122, USA }

\author{Antonio Suma} 
\affiliation{Dipartimento di Fisica, Universit\`a degli Studi di Bari and INFN, Sezione  di Bari, via Amendola 173, 70126 Bari, Italy}
\affiliation{Institute for Computational Molecular Science, Temple University, Philadelphia, PA 19122, USA }

\author{Daniel Sigg} 
\affiliation{dPET, Spokane, WA,  USA}

\author{Giuseppe Gonnella} 
\affiliation{Dipartimento di Fisica, Universit\`a degli Studi di Bari and INFN, Sezione  di Bari, via Amendola 173, 70126 Bari, Italy}

\author{Vincenzo Carnevale}
\affiliation{Institute for Computational Molecular Science, Temple University, Philadelphia, PA 19122, USA }

\begin{abstract}
We characterize equilibrium properties and relaxation dynamics of a two-dimensional lattice containing, at each site, two particles connected by a double-well potential (dumbbell). Dumbbells are oriented in the orthogonal direction with respect to the lattice plane and interact with each other through a Lennard-Jones potential truncated at the nearest neighbor distance. We show that the system's equilibrium properties are accurately described by a two-dimensional Ising model with an appropriate coupling constant.  Moreover, we characterize the coarsening kinetics by calculating the cluster size as a function of time and compare the results with Monte Carlo simulations based on Glauber or reactive dynamics rate constants. 
\end{abstract}

\maketitle

\noindent

\section{Introduction}

Many complex system can be described as a collection of interacting diffusing agents  with internal 
degrees of freedom associated with a discrete set of states.  Examples range from bacterial communities, 
to layers of cells in tissue development, to biomolecular systems~\cite{Prindlenature,camley2017physical,lipid_chapter}, where the internal states can represent either emission of electric or chemical signals, differences in motility, or  conformational changes of the  agents structure.
Importantly, the internal state of the agent/particles may affect the type and/or strength of pairwise 
interactions. This results in non-trivial and often interesting statistical properties under 
equilibrium or out-of-equilibrium conditions. 

Notable examples of biomolecular systems showing these features are biological membranes. Cell membranes are complex
mixtures of amphiphilic molecules, namely lipids, which show a complex phase behavior~\cite{Schmid_2017,pantelopulos2018regimes,usery2017line,honerkamp2012experimental}. Of particular relevance
are two macroscopic phases: the liquid disordered, $L_d$, and liquid ordered, $L_o$, that co-exists in ternary mixtures containing 
approximately equal amounts of cholesterol, unsaturated and saturated lipids. At the de-mixing transition, while unsaturated 
lipids partition into a $L_d$ phase, saturated lipids mostly segregate in cholosterol-rich, hexatically-ordered ($L_o$) domains in which the hydrophobic tails are in the extended 
conformation and give rise to a thicker layer compared to the surrounding $L_d$ phase~\cite{katira2016pre}. Interesting and still 
partly unanswered questions then concern the role of integral membrane proteins, which may have different affinities for the two 
macroscopic phases: does the presence of proteins change the phase behavior of the membrane? Is their lateral distribution or 
their internal conformational transitions affected by fluctuations in lipid composition?  

To address these questions, it is customary to define a statistical field that describes locally the two macroscopic phases $L_o$ and $L_d$, mapping their two components onto the celebrated Ising model (see, e.g.~\cite{machta2018}) with Hamiltonian: 
\begin{equation}
    H=-\sum_{<i j>}J s_i s_j ,
    \label{Hising}
\end{equation}
where $s_i=\pm 1$ is the particle's $i$ spin, $J$ is the coupling constant ($J>0$ for ferromagnetic systems) 
and the sum runs through nearest neighbors spins. 

While extremely insightful, studies based on this mapping rely on two inherent assumptions: spins are rigidly arranged on
a lattice, and their total density is fixed. These turn out to be severe limitations when modeling lipid membrane for which the surface
area (and thus the density) fluctuates and that, as mentioned above, undergo a transition from an hexatically-ordered  to 
a disorder state (from $L_o$ to $L_d$). It is therefore useful to envision a model Hamiltonian for a collection of particles that, while retaining a  
two-state character for the internal degree of freedom, can diffuse in space and therefore assemble and
form aggregates of varying densities and/or degree of symmetry.

Here we model
lipids as pairs of particles (dumbbells) with two stable particle-particle bond lengths (thereby mimicking the extended and disordered
conformations of the lipid tails). The dumbbells are oriented along the direction perpendicular to the 
membrane plane (z) and can undergo Brownian diffusion in the (x,y)-plane. Thus, we neglect here fluctuations perpedicular to the surface and consider only a planar membrane. Importantly, while one of the particles of each dumbbell
is constrained to the z=0 plane, the other one can fluctuate between the two minima of a Ginzburg-Landau type potential that
enforces a quasi-two-state behavior. We will call it the bond part of the potential (b), as it bonds the two particles of each dumbbell, and it has the general form:
\begin{equation}
V^b(r) = \epsilon\Big [a(r - c)^4 - b(r - c)^2\Big ],
\end{equation}
with a, b, c parameters and $\epsilon$ the characteristic system's energy. The particles of the dumbbells which are not constrained in the $z=0$ direction can also interact with their nearest neighbors with an attractive pairwise Lennard-Jones potential (p) of the form:
\begin{equation}
V^{p}(r) = 4\epsilon\Big [\bigg(\frac{\sigma}{r}\bigg)^{12} - \bigg(\frac{\sigma}{r}\bigg)^6\Big ],
\end{equation}
with $\sigma$ the typical system length. This potential represents the spin-spin interaction of Eq.~\ref{Hising}. More details will be provided in the Model section. The resulting Hamiltonian shows resemblances with previously described models, analyzed in the context
of stochastic resonance between anharmonic oscillators~\cite{Pikovsky2002}.

As a first steps toward modeling a collection of two-state Brownian particles, here we characterize the equilibrium and relaxation dynamics in the solid crystalline phase. In section~\ref{Sec2} of this paper, we describe in detail the model and numerical methods used throughout the paper. In the remaining part of the manuscript, we focus on characterizing the ferromagnetic properties of the model by considering a triangular lattice~\cite{potts1952spontaneous,Stephenson1964}, which represents the maximum packing configuration for Brownian particles in two dimensions. In section~\ref{Sec3}, we characterize the system's phase diagram and heat capacity. We also perform finite size scaling analysis to confirm that the critical exponents are the ones of the two-dimensional Ising model. In section~\ref{dynamical}, we characterize the single particle dynamics and the kinetics of clusters, comparing molecular dynamics with  Ising model simulations using Monte Carlo methods. Future work will extend this model to the off-lattice case to investigate how aggregation of Lennard-Jones particles couples with the order/disorder Ising phase transition. Importantly, we will then consider the effect of membrane proteins modeled as inclusions with fixed or variable height.

\begin{figure*}
\centering 
\includegraphics[width=1.9\columnwidth]{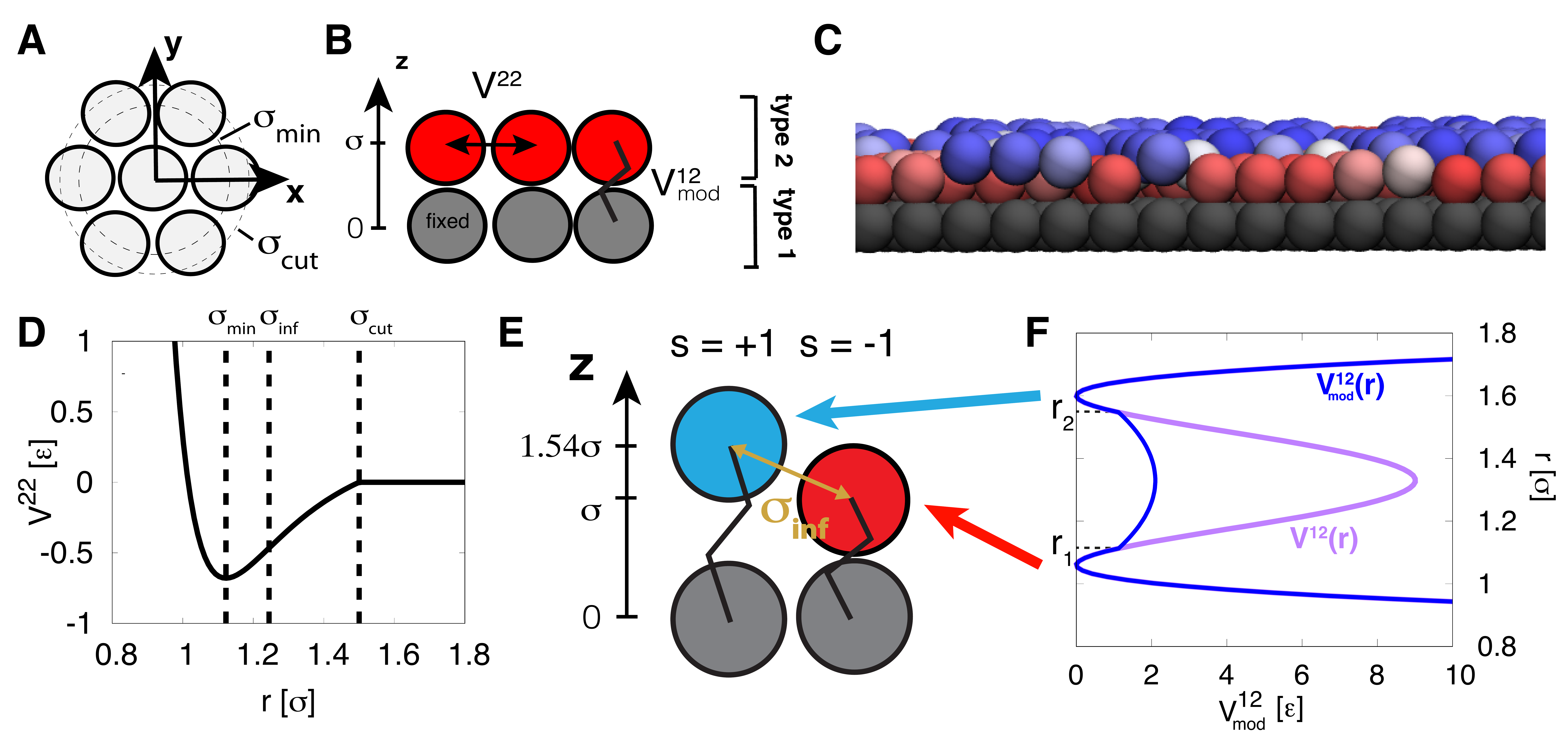}   
\caption{A) Top view of the system (on the x-y plane), showing the triangular lattice, along with $\sigma_{cut}=1.5\sigma$; $\sigma$ is the particles diameter and $\sigma_{min}=2^{\frac{1}{6}}\sigma$. B) Side view of the system (x-z plane). Each dumbbell consists of a bottom particle (type 1, grey) and a top particle (type 2, red) whose $z$ coordinate varies. The former are fixed in position, while the latter interact with potential $V^{22}$. Particles within each dumbbell interact through the potential $V_{mod}^{12}$. C) Instantaneous configuration of the dumbbells obtained by using the VMD software \cite{vmd}. D) Potential $V^{22}(r)$, with $\sigma_{cut}$, $\sigma_{min}$ and the inflection point $\sigma_{inf}=(\frac{26}{7})^{\frac{1}{6}}\sigma$ highlighted. E) Side view (on the x-z plane) of two interacting dumbbells in two different minima of $V_{mod}^{12}$. The minima are located at $z=\sigma$ (spin $s=-1$) and $z=1.54\sigma$ ($s=+1$). The type 2 particles are at a distance of $\sigma_{inf}$. F) Dumbbell  potential $V^{12}$ (purple), and its modified version $V^{12}_{mod}$ (blue) with the quadratic function used between $r_1=1.0496\sigma$ and $r_2=1.4904\sigma$. The two minima correspond to different $z$ values of type 2 particle.
}
\label{fig:model_sideview}
\end{figure*}

\section{Model and numerical methods}\label{Sec2}

\subsection{Model}\label{model}

We consider a two-dimensional triangular lattice composed of $N_d$ pairs of particles (dumbbells), for a total of $N=2N_d$ particles, oriented along the direction orthogonal to the lattice plane (z). The primitive vectors of the triangular lattice are $\bm{n}_1=[\sigma_{min},0,0]$ and $\bm{n}_2=[\frac{1}{2}\sigma_{min},\frac{\sqrt{3}}{2}\sigma_{min},0]$, where $\sigma_{min}=2^{\frac{1}{6}}\sigma$ is the minimum of the Lennard-Jones potential described below. This lattice implies 6 neighbors per dumbbell (Fig.\ref{fig:model_sideview}A).

The first layer of particles composing a dumbbell (type 1 particles), with Cartesian coordinates $\bm{r}_{1i}=[x,y,z]$ ($i=1,\ldots N_d$), is constrained on the plane $z=0$, while particles in the second layer (type 2 particles), with coordinates $\bm{r}_{2i}$  ($i=1,\ldots N_d$), can move along $z$, see Fig.\ref{fig:model_sideview}B-C. The $x$ and $y$ coordinates of all particles are fixed. Type 2 particles  interact with particles of the same type with a Lennard-Jones potential:
\begin{equation}
    V^{22}(r) =\begin{cases}
     4\epsilon((\frac{\sigma}{r})^{12} - (\frac{\sigma}{r})^6) - \delta_1 & 0 \le \sigma_{cut} \\
     0 & x > \sigma_{cut},
 \end{cases}
\end{equation}
where $\sigma_{cut}=1.5\sigma$ is the cutoff distance, and $\delta_1=4\epsilon((\frac{\sigma}{\sigma_{cut}})^{12} - (\frac{\sigma}{\sigma_{cut}})^6)$ ensures that the potential is continuous at $\sigma_{cut}$. The potential is thus attractive between the minimum $\sigma_{min}$ and the cutoff, and repulsive for $r<\sigma_{min}$ (Fig.\ref{fig:model_sideview}D). Since the coordinates of type 1 particles are fixed over time, no interaction potential is defined  between them. When dumbbells are allowed to diffuse in the x-y plane (a case that we do not discuss in this paper) an interaction potential analogous in form to $V^{22}$ is introduced also between particles of type 1.

Particles composing a dumbbell, $\bm{r}_{1i}$ and $\bm{r}_{2i}$, are connected by a quartic bond potential. This functional form results in two energy minima, which can be approximately mapped onto the spin states of an Ising model (Fig.\ref{fig:model_sideview}E).  In particular, the bond potential is defined as:
\begin{equation}
    V^{12}(r) = \epsilon\Big [a(r - c)^4 - b(r - c)^2\Big ]+\delta_2
\end{equation}
where $a, b$ are positive control parameters. The two potential minima are positioned at $r_{min}=c\pm \sqrt{\frac{b}{2a}}$, the barrier height is $h=\frac{b^2}{4a}$ and the constant $\delta_2=\epsilon\frac{b^2}{4a}$ conveniently sets the minimum of the potential to zero (Fig.\ref{fig:model_sideview}F). 
We choose $a=1699.2$ and $b=247.5$, such that the distance between the two minima is $0.54\sigma$ and the barrier is $h=9\epsilon$;  under these conditions, the two potential wells are sufficiently narrow and well separated that a meaningful mapping can be established between the $z$ coordinate of the particle and the two discrete spin states.  The constant $c=1.27\sigma$, ensures that the two minima are at $r=1\sigma$ and $r=1.54\sigma$ (Fig.~\ref{fig:model_sideview}F). 
This specific choice of distances ensures that, when two neighboring particles of type 2 are in different minima, their interaction energy corresponds to the Lennard-Jones potential at its inflection point (Fig.~\ref{fig:model_sideview}D-E), located at $\sigma_{inf}= (\frac{26}{7})^{\frac{1}{6}}=1.2444\sigma$, {\it i.e.} where the attractive force is at its maximum. 

For practical reasons, in numerical simulations we considered a modified version of $V^{12}(r)$ that greatly increases the rate of transitions between the two spin states. We truncated the potential around the local maximum, between the two values of $r$ corresponding to an energy of $\epsilon$ ($r_1=1.0496\sigma$ and $r_2=1.4904\sigma$). We then defined a quadratic potential in this region so that an energy barrier of $\epsilon$ would be present (the total barrier height from the minima is thus $2\epsilon$). The overall piecewise potential reads: 
\begin{equation}
V_{mod}^{12}(r) = \begin{cases}
     -\epsilon d(r-c)^2 + 2\epsilon &  r_1 \leq r \leq r_2 \\
      \epsilon [a(r - c)^4 - b(r - c)^2]+\delta_2 & otherwise 
    \end{cases}
\end{equation}
where $d = 1/(r_1-c)^2$. Note that the choice of $V_{mod}^{12}$, which is continuous at $r_1$ and $r_2$ (Fig.~\ref{fig:model_sideview}F), is a compromise between the contrasting requirements of having a surmountable energy barrier at thermal equilibrium and  enforcing a negligible occupancy of the intermediate states (to obtain an effective two-state behavior).  

In conclusion, the $z$ position of type 2 particles can be mapped onto a spin variable, with $z=1\sigma$ representing spin $s=-1$ and $z=1.54\sigma$ spin $s=+1$. The energy maximum is located at $z=1.27\sigma$ with a barrier of height $2\epsilon$. The coordinate $z_i$ of a dumbbell can thus be conveniently remapped into a continuous spin variable:
\begin{equation}
\phi_i=\frac{z_i-c}{\sqrt{b/2a}}
\end{equation}
with $\phi_i=\pm 1$ on the two energy minima. 

The time evolution of the system is analyzed by integrating an equation of motion of the Langevin type:
\begin{equation}
  m \ddot{\bm{r}}_{2i}=-\gamma \dot{\bm{r}}_{2i} -\frac{\partial V_{mod}^{12}}{\partial \bm{r}^{12}_{i}} - \sum_{i \neq j} \frac{\partial V^{22}}{\partial \bm{r}^{22}_{ij}}+\sqrt{2k_BT\gamma}\bm{\eta}_i(t),  
\end{equation}
where $\bm{r}^{22}_{ij}={\bm{r}}_{2i}-{\bm{r}}_{2j}$, $\bm{r}^{12}_{i}=\bm{r}_{2i}-\bm{r}_{1i}$.  The sum is referred to the dumbbell index. $T$ and $\gamma$  are the temperature and friction, respectively, of the thermal bath in contact with the system, $m$ the particle's mass, $\bm{\eta}_i$ is an uncorrelated Gaussian noise with zero mean and unit variance. $\gamma$ is set to 0.5, which corresponds to an intermediate  regime of friction (see Sec.~\ref{dynamical}), in order for spin-flip to occur efficiently over time.

\subsection{Simulations setup and numerical methods}
The potentials and the equation of motion described in Sec.~\ref{model} were implemented in a modified version of the LAMMPS software~\cite{Plimpton}. 
The reference units are the standard Lennard-Jones reduced units $m$, $\sigma$ and $\epsilon$, all set to unity, as well as $k_B=1$. Thus, the time units are $\tau_{LJ}=\sqrt{\frac{m\sigma^2}{\epsilon}}$. We considered temperatures in the range $T\in[0.1,0.5]$. The standard Velocity-Verlet algorithm was used to integrate the equation of motions, while the Langevin forces are implemented in LAMMPS following ref.~\cite{schneider1978molecular}.

For the purpose of sampling the phase space, we simulated a 30 by 30 particle lattice with periodic boundary conditions, for a total  $N_d=900$ dumbbells. The $x,y,z$ positions of type 1 particles were initialized at the lattice sites as prescribed in the previous section, while all type 2 particles were initialized to $z=1$ (spin -1), and then allowed to equilibrate. The same setup was used for the finite size scaling analysis with, systems of increasing number of particles $N_d=100, 400, 900, 1600, 10000$.

The dynamics of single spin-flip, described and characterized in Sec.~\ref{dynamical1}, was studied in a 6 by 6 lattice (36 dumbbells), and the small cluster spin-flip dynamics (Sec.~\ref{Scfd}) was studied in a system at least twice as large as the cluster-side.

For the purpose of studying the cluster growth (Sec.~\ref{dynamical3}), we simulated a 1024 by 1024 system for a total of $N_d=1024^2$ dumbbells. Type 1 particles were initialized as before, while type 2 particles had the $z$ position randomly assigned to the $z=1$ or $z=1.54$ positions. 

In order to efficiently explore the phase space of the system as a function of the temperature, we used also the metadynamics method~\cite{Laio12562}, similarly to ref.~\cite{Oh}, implemented in LAMMPS through the Colvars package~\cite{fiorin13}. The method uses non-Markovian dynamics to discourage the system from visiting repeatedly the same regions of the configuration space and it is therefore useful to sample configurations with large free energy.
In practice, a collective variable $v$ is used to coarse-grain the configuration space and a short-ranged (Gaussian) repulsive potential, $V(v(t))$, centered on the instantaneous value of the collective variable is added at fixed time-intervals to the unperturbed Hamiltonian of the system. It can be shown that the time integral of the biasing potential converges asymptotically to the underlying free energy surface~\cite{dama2014well}.
The collective variable of choice here is the average magnetization per dumbbell, defined as 

\begin{equation}
    M=\frac{1}{N_d}\sum_i \phi_i,
\end{equation} 
where the sum over $i$ is referred to the dumbbell index, and the magnetization distribution is concentrated in the interval $M\in[-1,1]$ (altough, strictly, $M$ is unbound). This rescaling allowed for easier comparison to the properties of the Ising model. 
Gaussians potentials were added every 1,000 timesteps for all temperatures, while the Gaussian height was set to $0.005\epsilon$ and the Gaussian width to 0.001. The latter was chosen to be smaller than the collective variable standard deviation at all temperatures.
The potentials of mean force (pmfs) were computed by averaging the final third of the trajectory, sampled every $10^5$ steps.

Metadynamics simulations were run for a maximum of $5\cdot 10^8$ timesteps. Simulations to compute the average magnetization, the heat-capacity and the finite size scaling were run without metadynamics for a maximum of $4\cdot 10^9$ timesteps. Simulations for sampling single spin events lasted $10^9$ timesteps. The largest system used to characterize cluster growth was run for about $2\cdot 10^7$ timesteps.

\subsection{Kramers formula for rate constants}
In Sec.~\ref{dynamical1}, we compared the molecular dynamics (MD) spin flip kinetics with the Kramers' intermediate-friction formula~\cite{KRAMERS1940}, which approximates the rate of escape of a Brownian particle from a free energy well. The approximate rate reads: 
\begin{equation}
    k=\frac{\omega_R}{2\pi\omega_b}\left(-\frac{\gamma}{2m}+\sqrt{\bigg (\frac{\gamma}{2m}\bigg )^2+\omega_b^2}\right)\exp\left(-\frac{\Delta F}{T}\right),
    \label{Kramers}
\end{equation} 
where $\gamma$ is the friction coefficient and $F(M)$ is the pmf. Parameters derived from the pmf are the reactant well frequency $\omega_R = \sqrt{F''(M_R)/m}$, the barrier frequency $\omega_b = \sqrt{-F''(M_b)/m}$, and the barrier height $\Delta F = F(M_b)-F(M_R)$.  Eq.~\ref{Kramers} approaches the well-known large friction form for large $ \gamma > m\omega_b$.

\subsection{Dynamical Ising model}
To compare the kinetics of MD with that of the Ising model, we considered a two-dimensional triangular lattice of spins governed by a master equation using reactive or Glauber dynamics. Note that the model we refer to here is the standard Ising model, with Hamiltonian of Eq.~\ref{Hising} and spins taking up only two discrete values $s_i=\pm 1$. Reactive dynamics~\cite{sigg2020microcanonical} are characterized by an Arrhenius-type rate constant $k = \nu e^{-\Delta E/2T}$, whereas in Glauber dynamics~\cite{Glauber} flip rates are saturable, described by $k = \nu e^{-\Delta E/2T}/(e^{-\Delta E/2T}+e^{\Delta E/2T})$. The rate factor $\nu$ determines the  time scale of kinetics, while $\Delta E = 4J(3-u)$ is the change in system energy for a spin to flip from $-1$ to $+1$ with $u = 0,..,6$ neighboring spins in the $+1$ position. Rate constants for the reverse transition ($+1$ to $-1$) are obtained by changing the sign of $\Delta E$. The extended triangular lattice in the shape of a parallelogram can be mapped to the traditional square Ising grid with 2 diagonal interactions in addition to the usual 4 orthogonal interactions for a total of 6 neighboring spin interactions. Ising model simulations were performed in real time using the Gillespie algorithm for kinetic Monte Carlo (MC)~\cite{gillespie1977exact}, in which waiting times between spin flips are given by $-\ln r/\sum_{i}k_i$, where $r$ is a uniform random number between 0 and 1, and $\sum_{i}k_i$ is the sum of transition rates taken over all spins $i$. A second random number determines which spin is flipped weighted by $k_i/\sum_{i}k_i$. The Gillespie algorithm is a true kinetic Monte Carlo simulation with exponentially-distributed waiting times between transition events.

\subsection{Computing cluster size}\label{FFT}

In Sec.~\ref{dynamical3} we computed the clusters size through the structure factor. Here we describe the numerical implementation.

The spin were first discretized to -1 or 1 (using $z=c$ as mid point). The bounding square rectangle was changed into a parallelogram 
conforming to the skew-geometry of the primitive cell. For each frame, the triangular lattice was digitized onto a $\sqrt{N_d} \times \sqrt{N_d}$ square matrix, using the same mapping of the Monte Carlo simulations. The digitized matrix was used to compute fast Fourier transform (FFT), with discrete vectors in real and reciprocal space ranging from 1 to $\sqrt{N_d}$ and $-\sqrt{N_d}/2$ to $\sqrt{N_d}/2$, respectively, and consequently the structure factor, see Sec.~\ref{dynamical3}  for the definition. Note that after digitizing the triangular lattice, the radial distance in reciprocal space has to be computed as 
 \begin{equation}
     k=\sqrt{{k_1}^2+(1/3)(2{k_2}-{k_1})^2},
 \end{equation}
 with $k_i$ the coordinate in reciprocal space, in order to conform to the original Euclidean distance in the triangular lattice. The average cluster size $L(t)$  was obtained from:
 \begin{equation}
\frac{ L(t)}{\sqrt{N}{\sigma_{min}} }=\frac{\sum_{k} h(k,t)/n(k)}{\sum_{k} h(k,t)k/n(k) } \approx 2\pi \frac{\sum_{k} h(k,t)/n(k)}{\sum_{k} h(k,t) },
 \end{equation}
where we approximate $n(k)\approx 2\pi k$, $h(k)$ are the summed bin values of the structure factor in radial intervals $(k-0.5, k+0.5]$ and $n(k)$ are the bin counts.

\section{Equilibrium properties}\label{Sec3}

\begin{figure*}
\centering 
\includegraphics[width=0.7\textwidth]{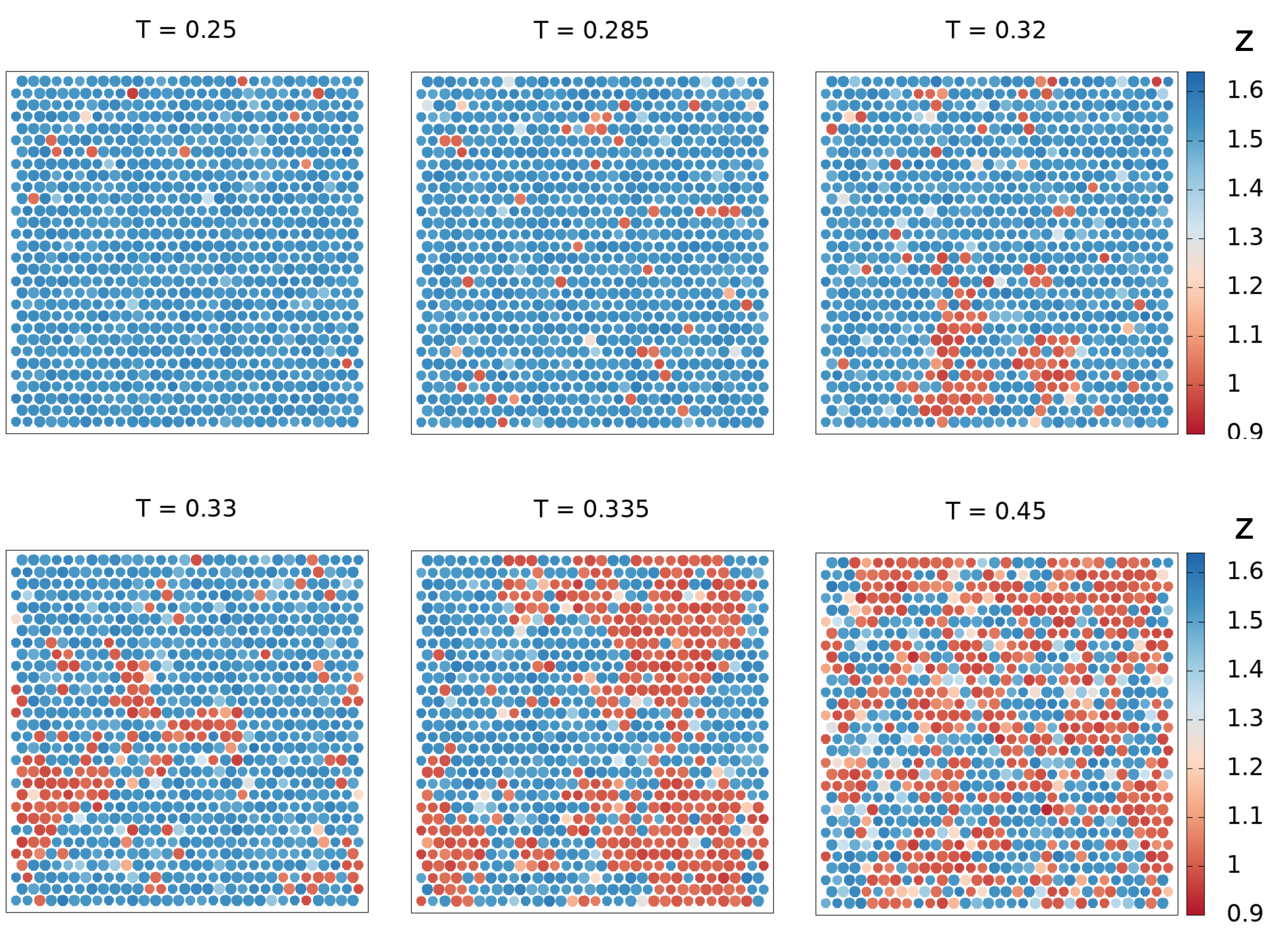}   
\caption{Equilibrium configurations sampled via molecular dynamics at six different temperatures, $T=0.25, 0.285, 0.32, 0.33, 0.335$. The system transitions from a magnetized state to a disordered one around $T= 0.335$.  Snapshot are taken from a top view in the x-y axis in which only type 2 particles are shown. Each particle is colored according its $z$ coordinate.}
\label{fig:equilibrium_snapshots}
\end{figure*}

Here we show that the properties of this collection of dumbbells, as defined in Sec.~\ref{Sec2}, closely match those of the Ising model. In particular, we will show how the magnetization and the heat capacity vary as a function of the temperature $T$.

\subsection{Phase diagram}

We performed metadynamics simulations of the system, for temperatures $T\in[0.25,0.39]$,
characterizing for each $T$ the free energy as a function of the average magnetization $M$, as described in Sec.~\ref{Sec2}. We found that, upon increasing the temperature, the system transitions from a fully ordered configuration in which all spins are aligned, to a configuration in which the average magnetization is zero. This is apparent from Fig.~\ref{fig:equilibrium_snapshots} showing the most probable conformations for six temperatures. At  $T=0.25, 0.285, 0.32, 0.33$ the conformations are magnetized and at $T=0.45$ the conformation is clearly disordered. Instantaneous configurations are rendered by showing type 2 particles from a top view colored according their $z$ coordinate.

\begin{figure}
\centering 
\includegraphics[width=1\columnwidth]{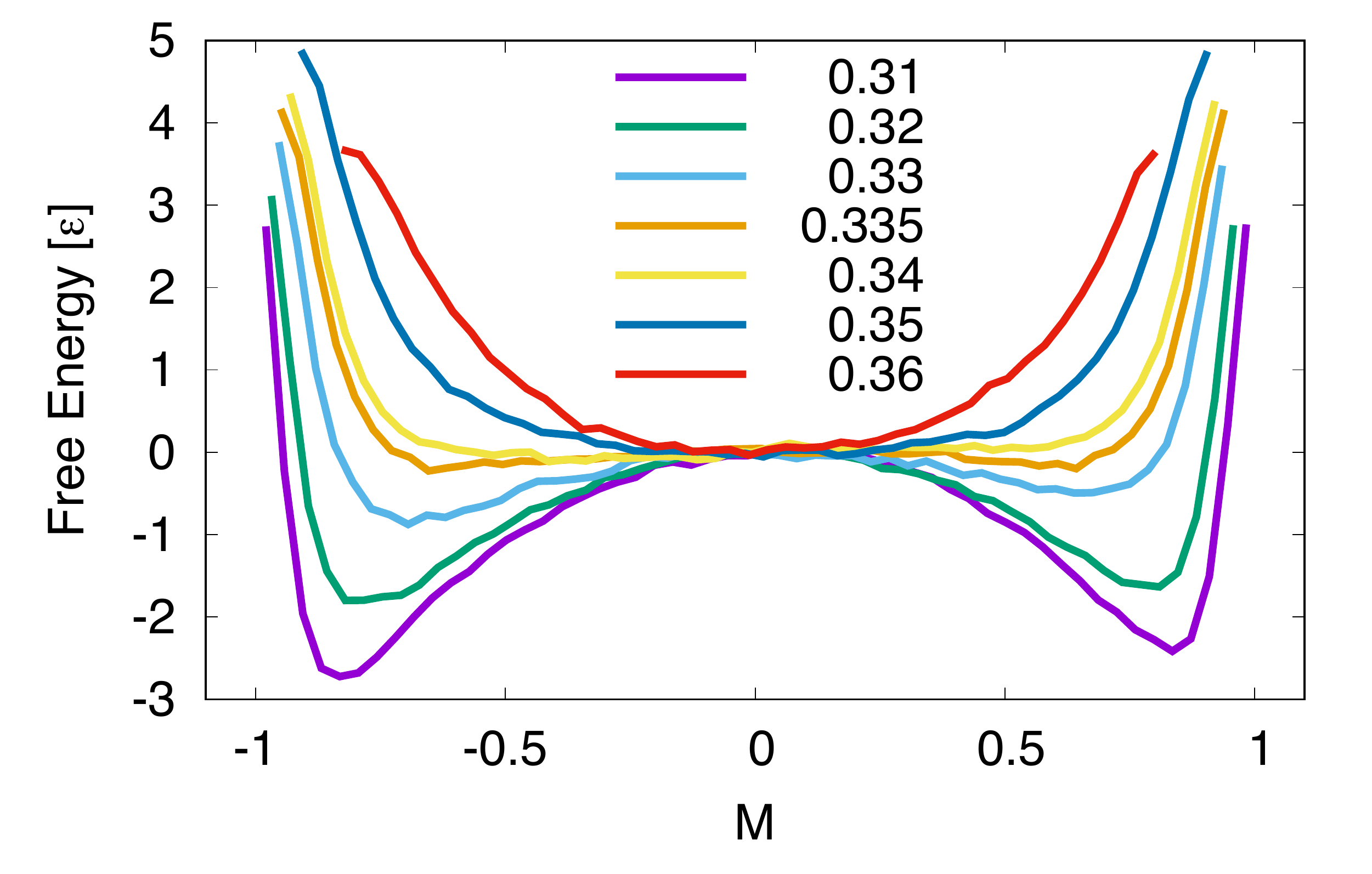}   
\caption{Free energy as a function of the average magnetization $M$ for 6 temperatures (see key), computed using metadynamics. Two distinct minima are visible with an energy barrier between them larger than $2k_BT$ up until $T = 0.33$.
}
\label{fig:pmf}
\end{figure}

Fig.~\ref{fig:pmf} shows the free energy as a function of $M$ for $T=0.31, 0.32, 0.33, 0.335,0.34, 0.35, 0.36$. Below $T=0.33$, the free energy presents two well-defined minima, separated by an energy barrier larger than $k_BT$. At $T=0.33$, the estimated free energy barrier between the two minima becomes of the order of $2k_BT$ allowing the system to explore the two oppositely aligned ferromagnetic states, a sign that the temperature is close to the critical one. By further increasing the temperature, the free energy profile flattens out completely and starts to develop a  single minimum at $M=0$, meaning that the system is in the disordered state. 

\begin{figure}
\centering 
\includegraphics[width=1\columnwidth]{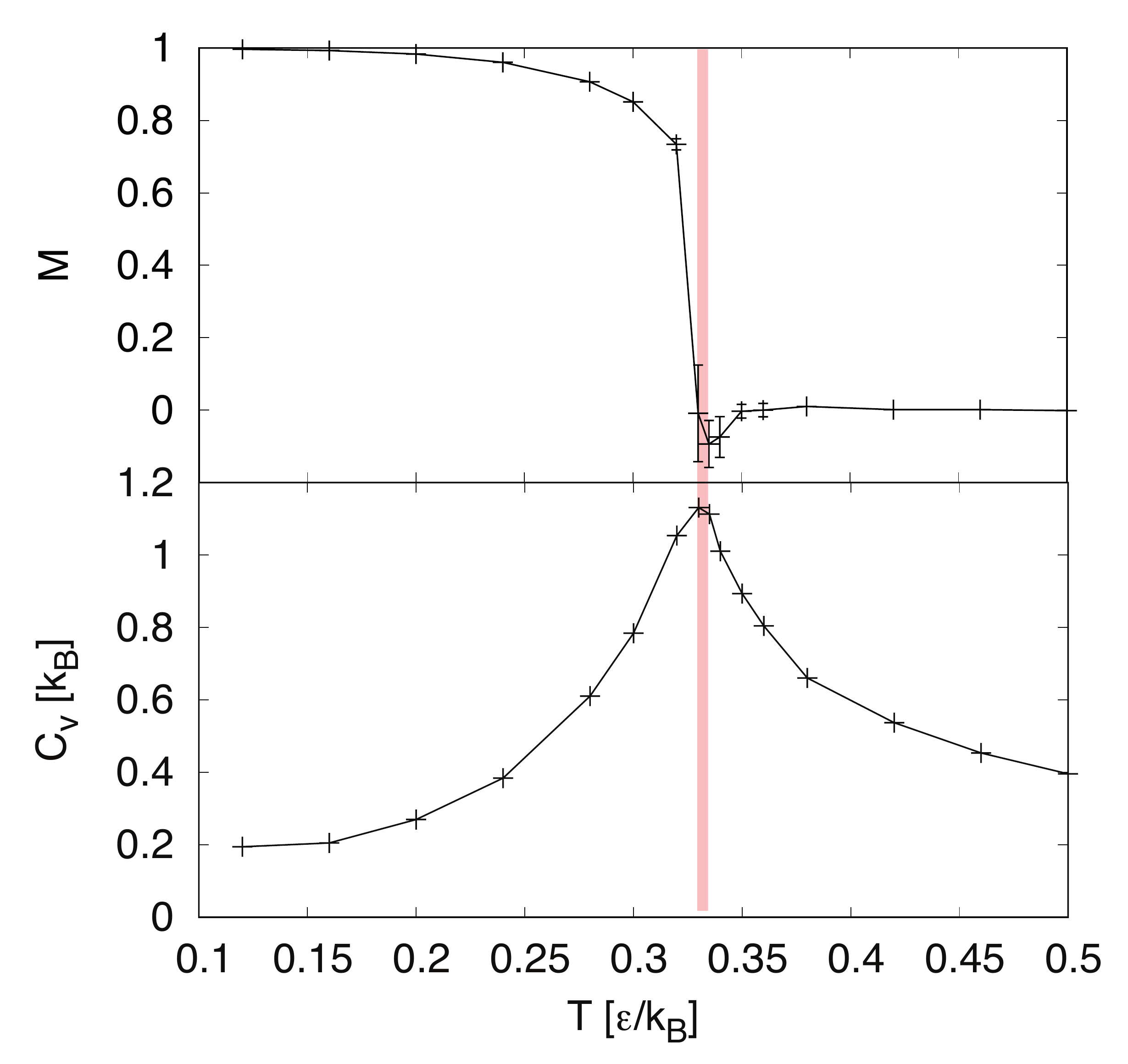}  
\caption{Magnetization and heat capacity as a function of the temperature, obtained from unbiased simulations. The red bar identifies the region where the critical temperature is located. 
}
\label{fig:mag}
\end{figure}

Fig.~\ref{fig:mag} shows the magnetization values measured from the positions of the free energy minima for temperatures below $T=0.33$ (red curve), as well as the average magnetization values obtained from long simulations performed without metadynamics. Consistent with the insight obtained from the free energy profiles, we observe that around $T\sim 0.33$ the magnetization quickly drops off to $M=0$. Deviations from this value at larger temperatures are due to limited sampling (see error bars).

\subsection{Heat capacity}

The heat capacity was calculated from trajectories sampled without the metadynamics technique, and thus unbiased, as  
\begin{equation}
C_{v} = N_d\frac{\langle E^2\rangle-\langle E \rangle^2}{k_BT^2},
\label{hceq}
\end{equation}
where $E$ is the average per-dumbbell potential energy. The  error bar on M was obtained using the blocking analysis~\cite{Flyvbjerg} in order to have uncorrelated data (the maximum decorrelation time observed is $10^4\tau_{LJ}$ ).

Results are shown in Fig.~\ref{fig:mag}.
Similar to the Ising model, a peak in the specific heat capacity is observed around $T\sim0.33$, suggesting a phase transition. 

\subsection{Finite size scaling for critical behavior}

\begin{figure*}
\centering 
\includegraphics[width=0.3\textwidth]{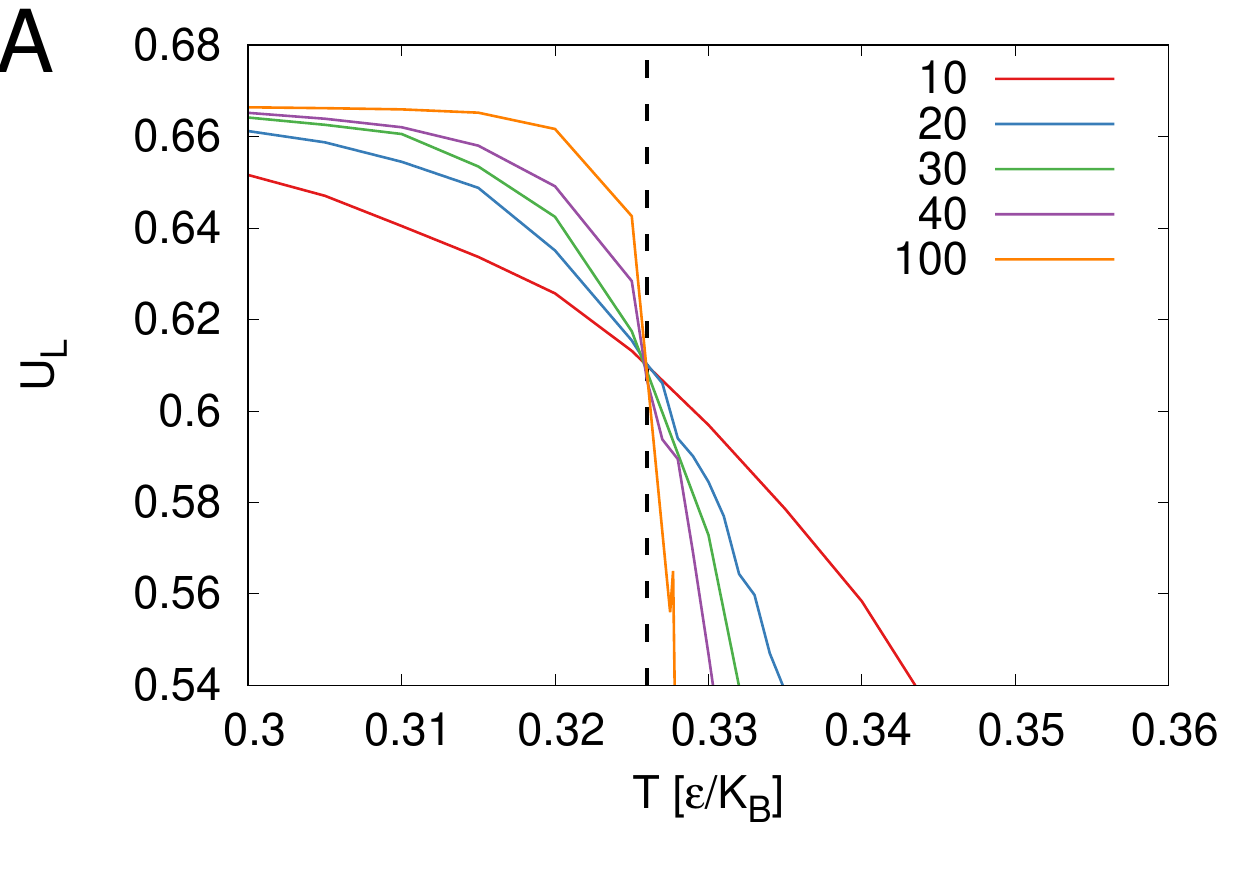}  
\includegraphics[width=0.3\textwidth]{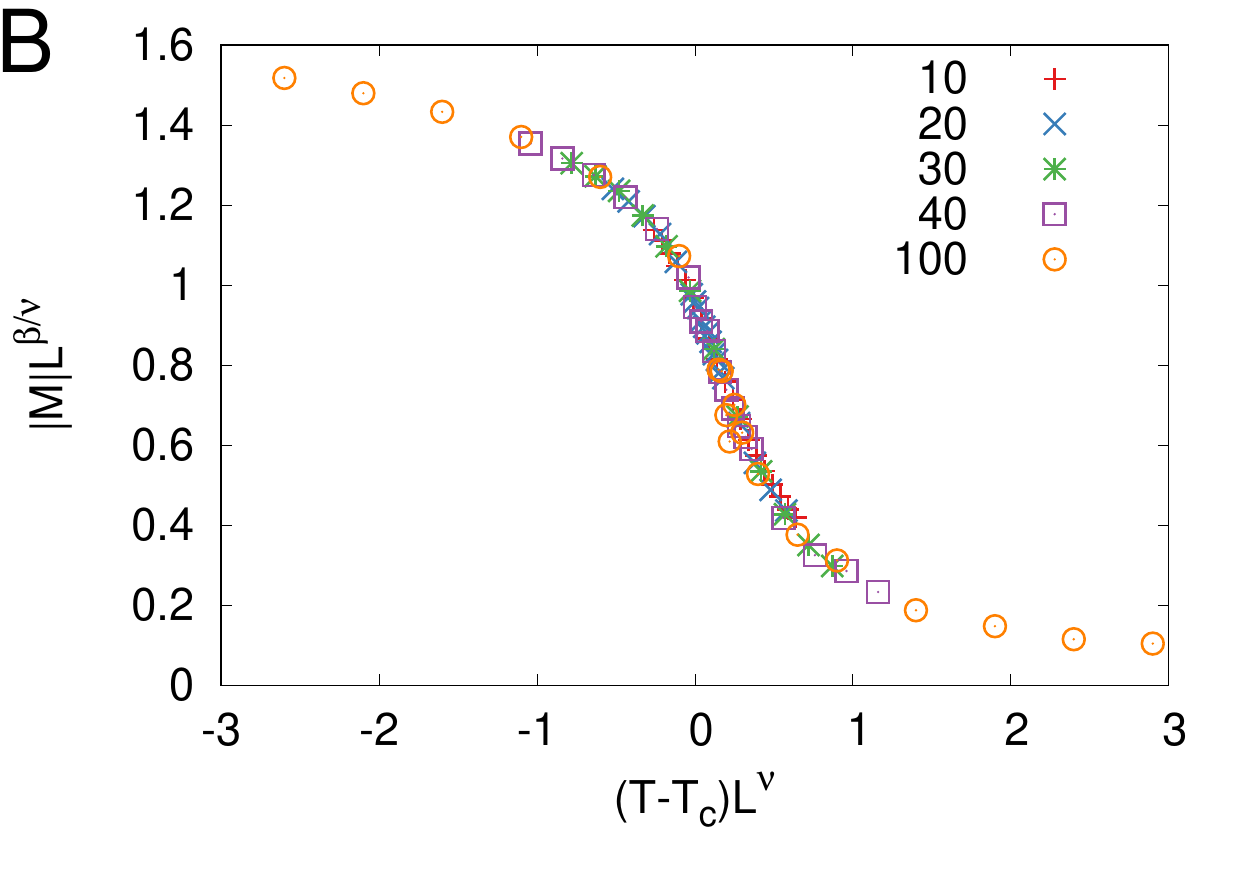}  
\includegraphics[width=0.3\textwidth]{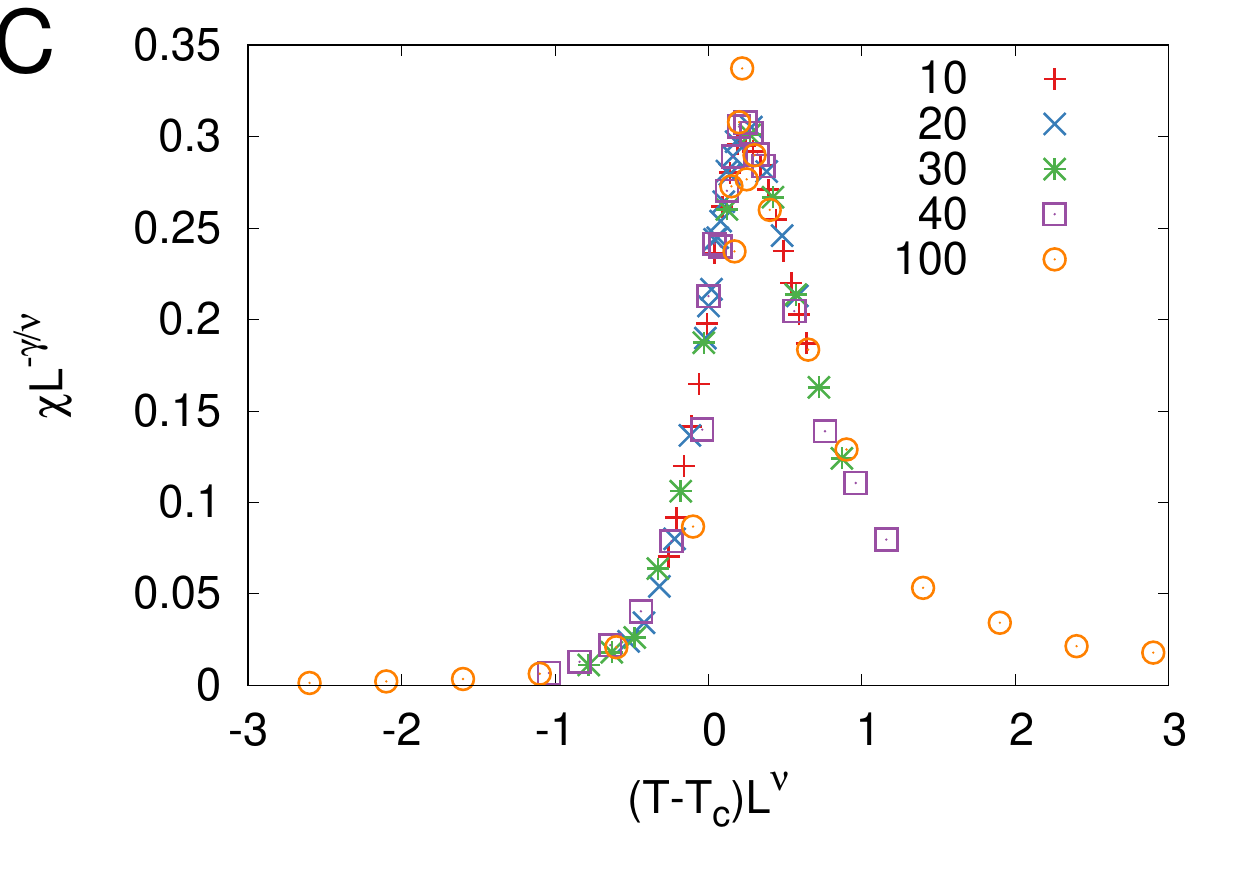}  \\
\includegraphics[width=0.3\textwidth]{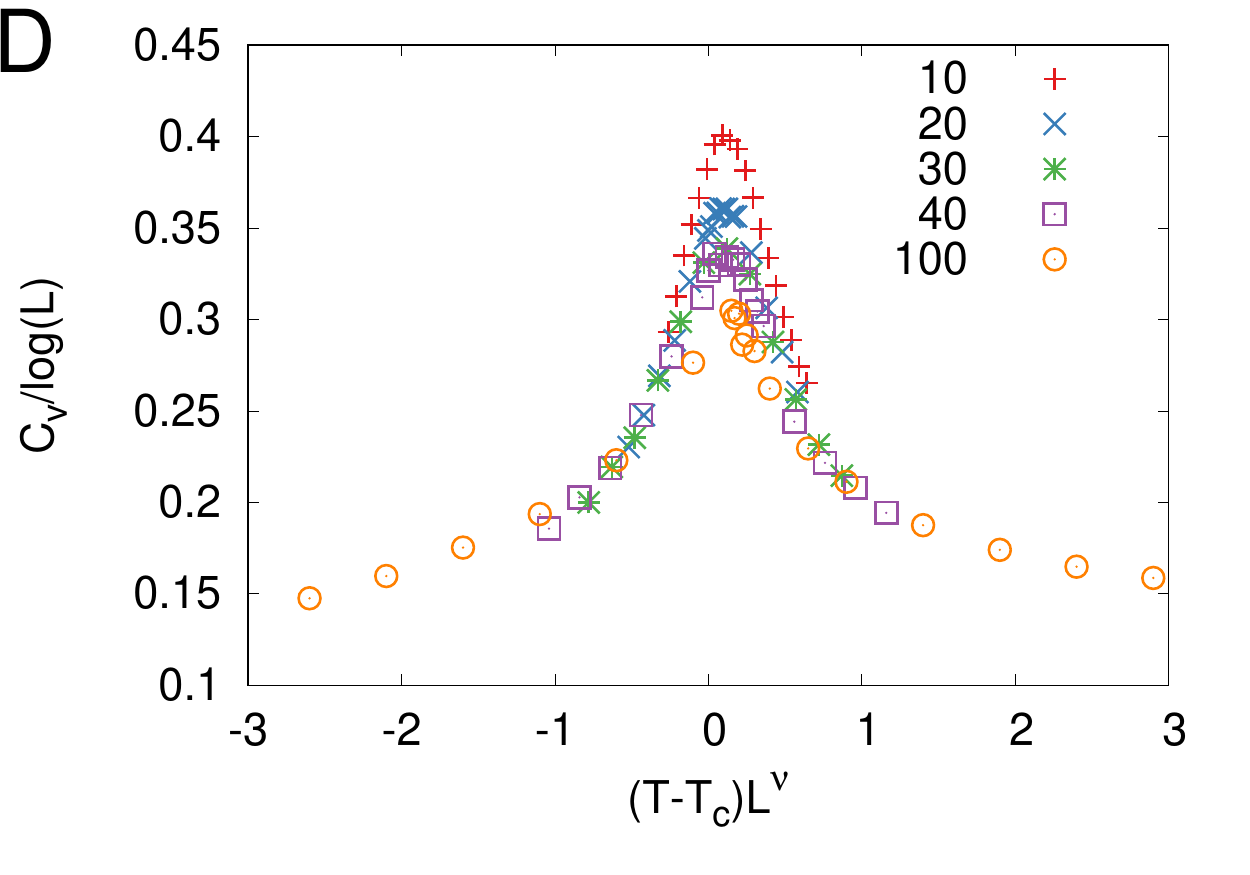}  
\includegraphics[width=0.3\textwidth]{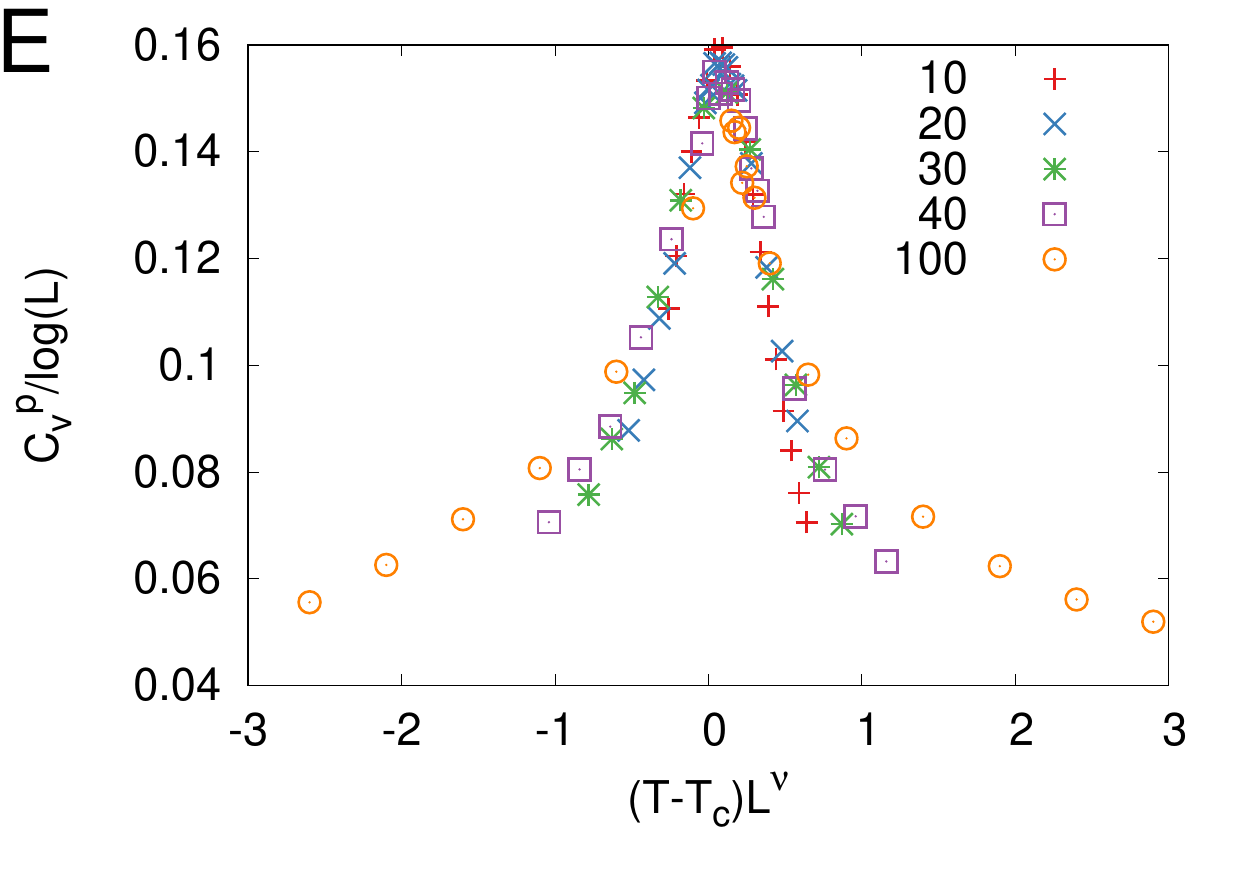}  
\includegraphics[width=0.3\textwidth]{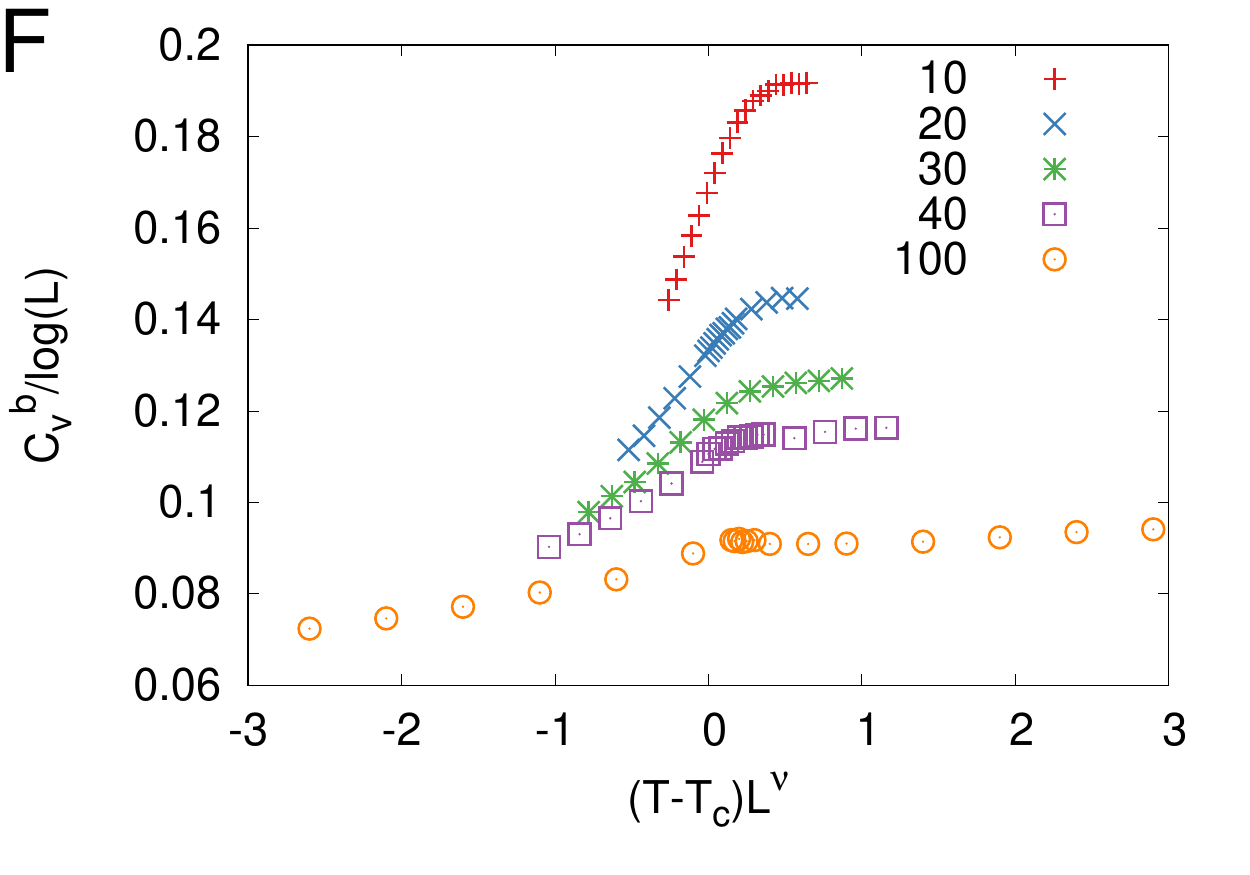}  \\
\caption{Finite size scaling analysis for system sizes L=10, 20, 30, 40, 100$\sigma$ (in the key). A) Binder cumulant, along with dashed line highlighting the crossing point between lines. Scaling of (B) absolute magnetization, (C)  susceptibility, (D)  heat capacity, (E)  heat capacity from pairwise interactions (p) and (F) from bonded ones (b). }

\label{fig:scaling}
\end{figure*}

We have shown that this model behaves as the Ising model. We now want to check if the critical behavior is compatible with the two-dimensional Ising universality class; thus we performed a finite-size scaling analysis~\cite{privman1990finite,cardy2012finite}. This analysis allows us to estimate the critical temperature by analysing the Binder cumulant, and to measure the critical exponent of the phase transition through the scaling of the absolute magnetization, susceptibility and heat capacity. We analyzed several system sizes L=10, 20, 30, 40, 100$\sigma$, with $N_d=100, 400, 900, 1600, 10000$ and temperatures near $T\sim0.33$.

The Binder cumulant is defined as~\cite{binder1981finite}:
\begin{equation}
U_L = 1-\frac{\langle M^4\rangle}{3\langle M^2\rangle^2}.
\end{equation}
The curves cross in Fig.\ref{fig:scaling}A at  $T_c\sim0.326$, which gives an estimate of the critical temperature. We will see on the next section, that this value agrees with the analytical solution for the triangular Ising lattice~\cite{Stephenson1964} .

We then proceed to rescale the thermodynamical quantities using $T_c\sim 0.326$ and the 2D Ising critical exponents $\nu=1, \beta=1/8, \gamma=7/8, \alpha=0$, which we assume here {\it a priori} to be the corrrect exponents. The first quantity we examine is the absolute magnetization $|M|$, defined as:
\begin{equation}
    |M|=\bigg |\frac{1}{N_d}\sum_i \phi_i\bigg |.
\end{equation} 
The correct scaling of this quantity is:
\begin{equation}
    |M|(T,L)=L^{-\beta/\nu}|\tilde{M}|(L^{1/\nu}(T-T_c))
\end{equation} 
 with $|\tilde{M}|$ independent on L.  Fig.\ref{fig:scaling}B confirms that the choice of critical exponents $\nu=1$ and $\beta=1/8$ collapses the data for all the system sizes into a single master curve.

The susceptibility is computed as
\begin{equation}
    \chi_{|M|}=N_d(\langle |M|^2\rangle-\langle |M|\rangle^2)
\end{equation} 
and scales as
\begin{equation}
    \chi_{|M|}(T,L)=L^{\gamma/\nu} \tilde\chi_{|M|}(L^{1/\nu}(T-T_c)),
\end{equation} 
with $\gamma=7/8$, as confirmed by the plot in Fig.\ref{fig:scaling}C.

The last quantity of interest is the heat capacity, see Eq.~\ref{hceq}. Here, as $\alpha=0$, the correct scaling should be logarithmic:
 \begin{equation}
    C_v(T,L)=\log(L) \tilde C_v(L^{1/\nu}(T-T_c)).
\end{equation} 
Note that, unlike the Ising model, here the potential energy function is composed by a pairwise interaction term, which represent the Lennard-Jones interactions and is directly related to Eq.~\ref{Hising}, and a bond term, implemented as a truncated quartic potential which represents the spring connecting the two beads of each dumbbell. In Fig.\ref{fig:scaling}D we can see the $C_v$ computed from the total potential energy, while Fig.\ref{fig:scaling}E-F show $C_v^{p}$ and $C_v^{b}$, computed, respectively, for the pairwise and bond interactions energy contributions. Interestingly, while $C_v^{p}$ scales correctly $C_v^{b}$ does not. The explanation is that fluctuations in the per-particle bond energy scale as $N_d^{-1}$ (since the bond energy for any given dumbbell does not depend on the system size), thus the specific $C_v^{b}$ is independent of $L$. For this reason, the relative contribution of $C_v^{b}$ to $C_v$ vanishes in the thermodynamic limit. In fact, we can see how $C_v^{b}/\log(L)$ decreases as $L$ increases (Fig.\ref{fig:scaling}F). Overall, the scaling of $C_v$ is expected to be the correct one for large enough $N_d$.

\section{Dynamical Properties}
\label{dynamical}

\begin{figure*}
\centering 
\includegraphics[width=1.8\columnwidth]{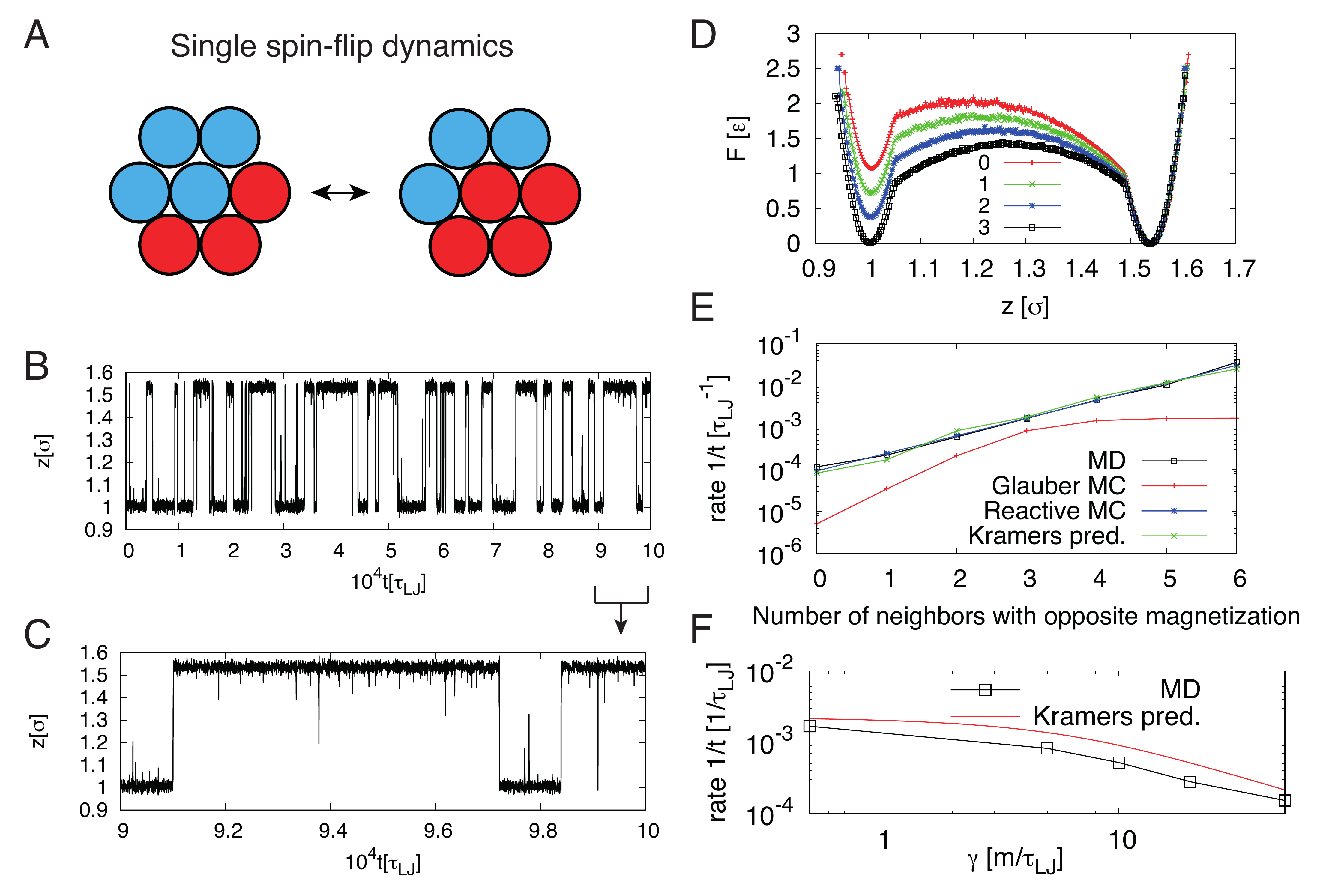}   
\caption{A) Illustration of the setup used to study single spin-flip dynamics. The central spin is allowed to flip, while the six neighbor spins have fixed spins, see text for detail.   B) Time evolution of $z$ for the central spin, with setup as in A). C) Zoom-in of B) over a smaller time interval. D) Free energy profile, as a function of $z$, with 0 to 3  neighbor spins with negative magnetization. E) Rate of central spin-flip with 0 to 6 neighbor spins with opposite magnetization, along with Glauber and reactive dynamics MC rates adjusted to fit the MD, see text. In green, Kramers prediction based on Eq.~\ref{Kramers}. F) Rate of central spin-flip with same setup as A), as function of $\gamma$. The temperature is fixed to $T=0.185$.
}
\label{fig:sfp}
\end{figure*}

\begin{figure}
\centering 
\includegraphics[width=\columnwidth]{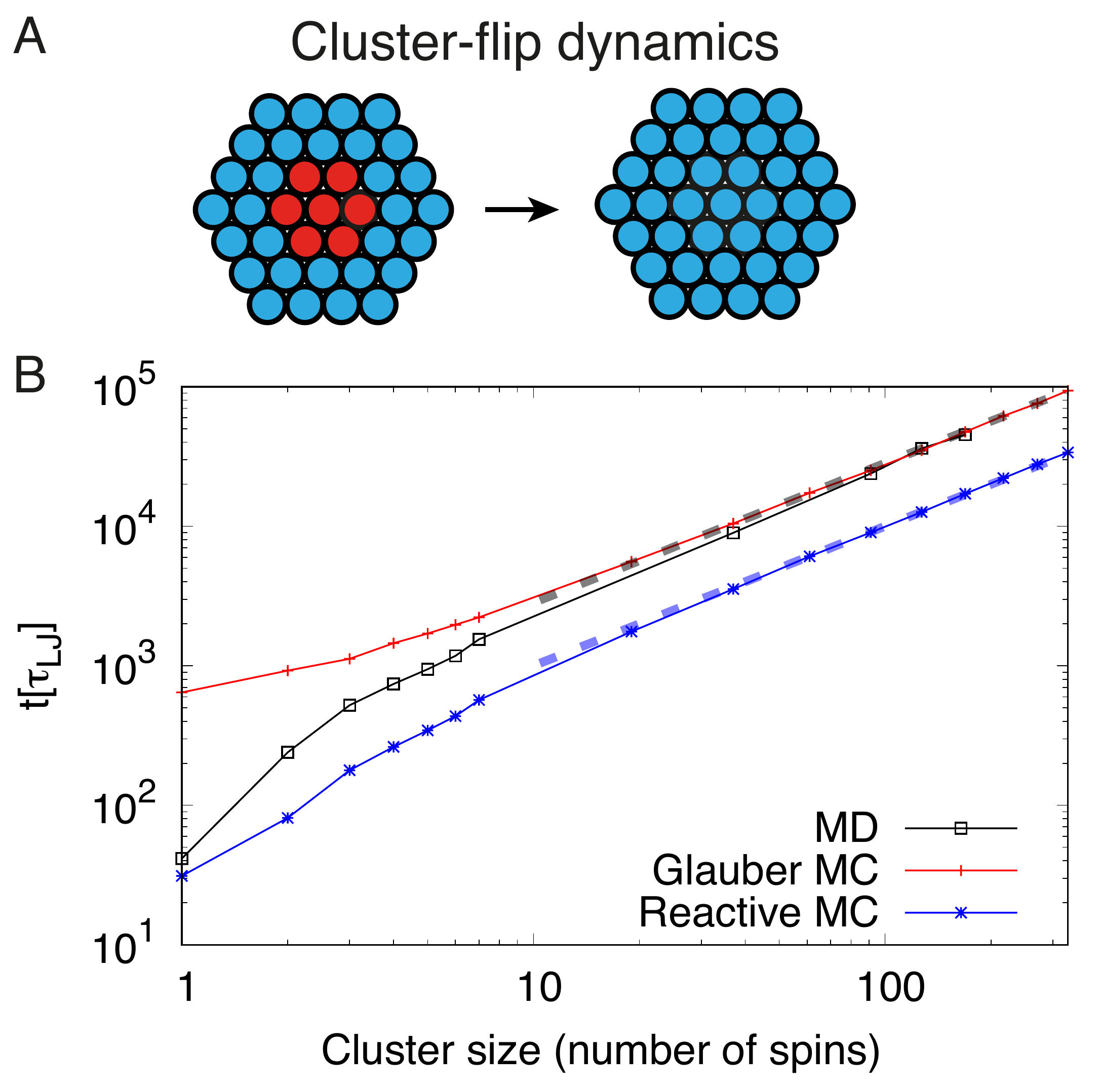}   
\caption{A) Illustration of the setup used to study cluster-flip dynamics. A single cluster is placed in the center of a system with opposite magnetization. B) Time required to invert the cluster's magnetization, as a function of the cluster size, in term of spins number, for  MD and reactive and Glauber MC.
}
\label{fig:csf}
\end{figure}

\begin{figure*}[h]
\centering 
\includegraphics[width=2\columnwidth]{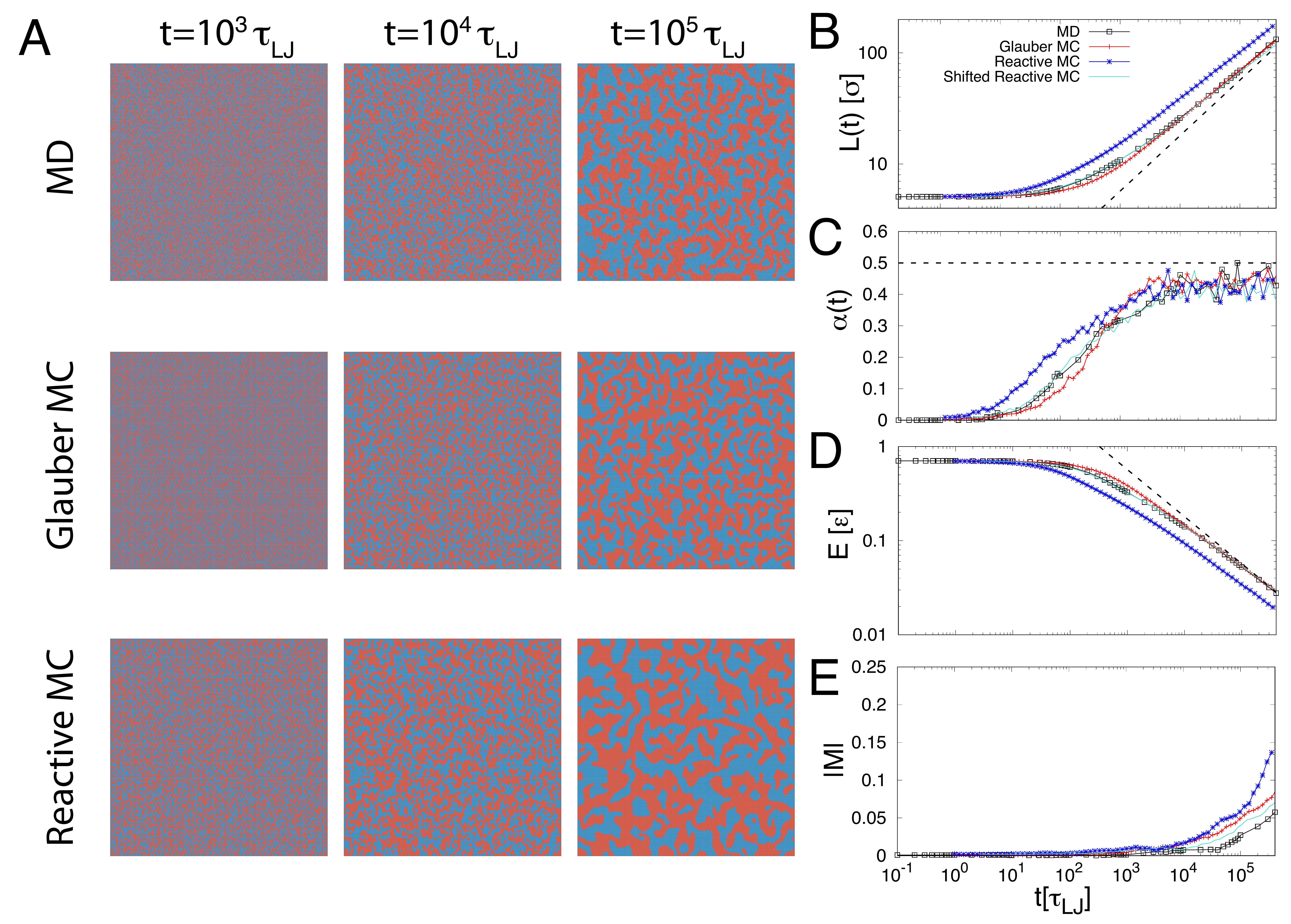}   
\caption{Cluster growth, starting from an initial random configuration at $T=0.185<T_c$, as a function of time. A) Three snapshots are shown at $t=10^3, 10^4, 10^5 \tau_{LJ}$ for MD simulations, and the corresponding snapshots for Glauber and reactive dynamics MC. B) Average cluster size $L(t)$, C) growth exponent $\alpha(t)$, D) Ising model energy and E) absolute value of magnetization $|M|$ as a function of time for the three different dynamics considered. In cyan, reactive dynamics MC shifted by a factor of $2.75$, see main text for details. In B) the dashed line shows the power law fit with exponent $0.5$. In D) it shows a power law with exponent $-0.5$.
}
\label{fig:structure_factor}
\end{figure*}

From a dynamical point of view, the trajectory obtained from integration of a Langevin-type equation of motion is conceptually distinct from the time series obtained from a Monte Carlo (MC) simulation. Thus it is interesting to investigate the dynamical differences between MC and molecular dynamics for these Ising-like systems.

\subsection{Single spin-flip dynamics}
\label{dynamical1}

First, we characterize the spin-flip dynamics for a single spin. In particular, we focus on cases where the six neighbor spins have their magnetization fixed, Fig~\ref{fig:sfp}A. This is obtained by constraining the positions to $z>c$ for $s=+1$ and $z<c$ for $s=-1$, using  for each spin a repulsive quadratic potential at $z=c$ with force constant $k=10\epsilon/\sigma^2$. Seven cases are possible, with a total of 0 to 6 neighbor spins with opposite magnetization to the central one that has to flip (and a complementary number of 6 to 0 neighbor spins with same magnetization as the central one). The temperature will be from now on fixed at $T=0.185$ for MD, which is about $0.56 T_c$ and we choose similarly $T/T_c=0.56$ for MC (we use in this case the theoretical value for $T_c$).

Fig~\ref{fig:sfp}B-C show a single spin-flip trajectory over time with three neighbors with same magnetization and three with opposite one. The value of $z$, and consequently of $M$, varies continuously in time. Fig~\ref{fig:sfp}D shows the free energy profile, as a function of $z$, of a spin with 0 to 3 neighbor spins with negative magnetization. The other cases have a  free energy which is the mirror image around $z=c$.

From the 0 neighbor free energy of Fig~\ref{fig:sfp}D, it is possible to estimate an effective coupling constant $J$ (see Eq.~\ref{Hising}) and thus infer the critical temperature $T_c$. Because we have six nearest neighbors, $2J=\Delta E/6$, with $\Delta E=1.074\epsilon$ the difference in energy between the two minima. The value of  $J=0.0895$ can be then inserted in the analytical solution for the triangular Ising lattice~\cite{Stephenson1964} $\frac{k_BT_c^{th}}{J} = 3.640957$, finding $T_c^{th}=0.326$. This estimate is in striking agreement with the one obtained using the Binder cumulant, thereby further confirming the presence of a phase transition in the thermodynamic limit. Note that the value of $J$ extracted from the free energy is different from the one obtained by the bare difference in interaction energy between the ``parallel'' and ``anti-parallel'' case, which amounts to $2J=[V^{22}(\sigma_{inf})-V^{22}(\sigma_{min})]=0.202$. At the same time, we expect a slight dependence of the value of $T_c$ by changing the bond potential around the barrier (and consequently $J$).

Fig.~\ref{fig:sfp}E shows the rate (inverse time) for the central spin to invert its magnetization, for the seven possible cases. The flipping-rate increases exponentially, as expected due to the linear decrease in the energy barrier observed in Fig~\ref{fig:sfp}D. This results are in agreement with the prediction obtained from Kramers' intermediate-friction rate constant formula (Eq.~\ref{Kramers}) using MD-derived pmfs to compute the formula parameters. 

Note that changing the barrier height of the bond potential by a factor $\alpha$ affects the typical crossing time from one state to the other by a factor $e^\alpha$: this exponential scaling makes it extremely difficult to sample events for large values of the barrier using MD simulations, unlike Monte Carlo methods. 

Fig. ~\ref{fig:sfp}F shows how the flipping rate with 3 neighbor spins of opposite magnetization is affected with changing $\gamma$, along with Eq.~\ref{Kramers}. Indeed, the chosen value of $\gamma = 0.5$ falls within the intermediate-friction regime.

We can use the MD results of Fig.~\ref{fig:sfp}E to match the rate constants of the kinetic Ising model. By varying $\nu$, $J$, and $T$ with arbitrary units, the Ising rate constants for the reactive dynamics can be aligned with MD-derived flip rates (Fig. ~\ref{fig:sfp}E). The Glauber rates, obtained using the same $\nu$ as the  reactive dynamics, underestimate reaction rates for all conditions (Fig.~\ref{fig:sfp}E). The two rates can coincide for the specific case of equal number (3) of oppositely magnetized neighbor particles by scaling Glauber rates by a factor of 2.

\subsection{Cluster-flip dynamics}\label{Scfd}

Next, we characterized the amount of time needed for an hexagonal cluster to completely invert its magnetization when surrounded by a bulk of spins with opposite magnetization  (Fig.~\ref{fig:csf}A). Note that only spins inside the cluster were considered when monitoring the magnetization.

Fig.\ref{fig:csf}B reports on the cluster-flip time. The time scales linearly with the cluster size for sufficiently large clusters, while a non-trivial dependency is observed for low cluster sizes, which is dependent on the chosen dynamics. Although the rates between MD and reactive dynamics were perfectly matched (the curves start from the same point) the cluster-flip times do not match for large clusters. The ratio between the two slopes is $2.75$. Instead, for Glauber MC dynamics these times coincide for large clusters.  

\subsection{Cluster growth kinetics}
\label{dynamical3}
 
Last, we characterize the time evolution of cluster size $L(t)$ by performing a similar analysis to the one reported in ref.~\cite{gonnella2014phase}. Fig.~\ref{fig:structure_factor}A shows the evolution of the $1024^2$ spin system starting from a random configuration for the three different dynamics considered.

We computed the structure factor in the reciprocal space defined by the contravariant vectors $\bm{b}_1=[\frac{1}{\sigma_{min}},-\frac{1}{\sqrt{3}\sigma_{min}},0]$ and $\bm{b}_2=[0,\frac{2}{\sqrt{3}\sigma_{min}},0]$, such that $\bm{n}_1\cdot\bm{b}_2=\bm{n}_2\cdot\bm{b}_1=0$. The structure factor is the spectral density of the spin matrix, where the spins have been considered in their discretized version $s$:
 \begin{equation}
     S_{F}(\bm{k})=|{\sum_{\bm{r}}(s(\bm{r})-\langle{s}\rangle)e^{-2\pi i\bm{k}\cdot\bm{r} }}|^2,
 \end{equation}
where $\bm{r}$ are the positions in real space ranging from $[1,\ldots,\sqrt{N_d}]\bm{n}_i$, $\bm{k}$ the ones in the reciprocal space, ranging between $[-\sqrt{N_d}/2,\ldots,\sqrt{N_d}/2](\bm{b}_i/\sqrt{N_d})$ and $\langle{s}\rangle$ is an average over all configuration's spins.  
 For each frame, we computed $S_{F}(\bm{k},t)$ and
the average cluster size $L(t)$  was obtained from:
 \begin{equation}
 L(t)=\frac{\int  \bm{dk} S_{F}(\bm{k},t)/k}{\int  \bm{dk} S_{F}(\bm{k},t)}.
 \end{equation}
For the actual numerical implementation details, see Sec.~\ref{FFT}. 

 Fig.~\ref{fig:structure_factor}B shows $L(t)$, along with a power law (dotted line) with exponent 0.5, which is expected when the order parameter is non-conserved as in model A dynamics \cite{bray2003coarsening}, see also Supplementary Movie S1 (system with $100^2$ spins, $T=0.185$ starting from a random configuration). Consistent with the observations of Sec.~\ref{Scfd}, $L(t)$ of MD and Glauber MC overlaps well at long times, while reactive dynamics MC does not. The latter overlaps when the times are rescaled by the factor $2.75$ (cyan curve), as measured in Sec.~\ref{Scfd}.  
 
Fig.~\ref{fig:structure_factor}C shows the growth exponent $\alpha$, $L(t)\sim t^{1/\alpha}$ as a function of time, computed as~\cite{corberi2008influence} 
 \begin{equation}
 \frac{1}{\alpha(t)}=\frac{d\ln L(t)}{d\ln t}.
 \end{equation}
The exponent slowly saturates to $0.5$ over time. This feature is common to both MD and MC, and it is due to the dynamics of cluster interfaces, which play an important role at finite ($T>0$) temperatures~\cite{corberi2008influence}. 

Finally, Fig.~\ref{fig:structure_factor}E-F show the system's energy $E(t)$ measured using the Ising model Hamiltonian~\ref{Hising}, after spin discretization for MD configurations, and the total system's magnetization $|M(t)|$. The former saturates to a  power law (dotted line) with exponent -0.5.

\section{Discussion}

In this work, we investigated a two-dimensional system of pairs of particles (dumbbells) connected by a double-well potential. This allows the dumbbells to fluctuate between two states, which can be interpreted as two spin states $s_i=\pm 1$. Importantly, dumbbells interact with each other through a Lennard-Jones potential with an interaction energy that discriminates between ``parallel''  and ``anti-parallel'' configurations. In this way, each spin can influence its closest neighbors (six for the triangular lattice considered here). All spins evolve simultaneously under the action of the interaction potential as prescribed by a Langevin-type equation of motion.

First, we characterized equilibrium properties, included a finite size scaling analysis: the observed behavior follows closely what expected for an Ising model, with the critical exponents matching those of the two-dimensional Ising universality class. Second, we characterized several dynamical properties, such as single spin- and cluster-flip, and the cluster growth kinetics starting from a random initial configuration. We highlighted a rich dynamical behavior, which differs to some extent from two of the most widely considered dynamics, namely reactive and Glauber dynamics Monte Carlo. 

Future work will investigate the most general case in which dumbbells are allowed to diffuse in the $x, y$ plane, as it occurs in biological systems such as lipid membranes. We expect the phase diagram of this off-lattice case to show a non-trivial combination of vapor-liquid and liquid-solid transitions typical of both Lennard-Jones systems and ferromagnetic order/disorder transitions.

{\it Acknowledgements}

This work was funded by the National Institutes of Health (R01GM093290, S10OD020095 and R01GM131048; V.C.), and the National Science Foundation through grant IOS-1934848 (V.C.). This research includes calculations carried out on Temple University's HPC resources and thus was supported in part by the National Science Foundation through major research instrumentation grant number 1625061 and by the US Army Research Laboratory under contract number W911NF-16-2-0189.


%

\end{document}